\documentclass{article}

\usepackage{arxiv}

\usepackage[utf8]{inputenc} 
\usepackage[T1]{fontenc}    
\usepackage{hyperref}       
\usepackage{url}            
\usepackage{booktabs}       
\usepackage{amsfonts}       
\usepackage{nicefrac}       
\usepackage{microtype}      
\usepackage{graphicx}
\usepackage[numbers]{natbib}
\usepackage{doi}
\usepackage{enumitem}
\usepackage{makecell}
\usepackage{titlesec}
\usepackage{amsmath}
\usepackage{caption}
\usepackage[table]{xcolor}
\setlist{leftmargin=3.0mm}

\titlespacing\subsection{0pt}{12pt plus 4pt minus 2pt}{0pt plus 4pt minus 2pt}

\title{Structure-Agnostic Prediction of the Electronic Density of States with a Chemical Language Model}

\author{
    \href{https://orcid.org/0009-0003-5578-990X}
    {\includegraphics[scale=0.07]{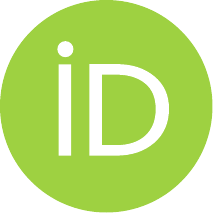}\hspace{1mm}Ivan D. Rubtsov}\\
    MSU Institute for Artificial Intelligence\\
	Lomonosov Moscow State University\\
	Moscow 119192, Russia\\
	\And
	\href{https://orcid.org/0009-0006-8637-0820}
    {\includegraphics[scale=0.07]{orcid.pdf}\hspace{1mm}Ivan V. Dudakov} \\
    MSU Institute for Artificial Intelligence\\
	Lomonosov Moscow State University\\
	Moscow 119192, Russia\\
	\And
	\href{https://orcid.org/0000-0001-6117-5662}
    {\includegraphics[scale=0.07]{orcid.pdf}\hspace{1mm}Vadim V. Korolev}\thanks{\textit{Email address}: \texttt{korolevvv\_01@my.msu.ru}}\\
    MSU Institute for Artificial Intelligence\\
	Lomonosov Moscow State University\\
	Moscow 119192, Russia\\
}

\renewcommand{\shorttitle}{Structure-Agnostic Prediction of the Electronic Density of States}

\begin{document}
\maketitle

\begin{abstract}
The electronic density of states (DOS) is conventionally computed from a relaxed crystal structure, which is unavailable for compounds that have been neither synthesized nor cataloged. Here we introduce DOSSIER (\textbf{D}ensity \textbf{o}f \textbf{S}tates from \textbf{S}to\textbf{i}chiometry with \textbf{E}ncoder \textbf{R}epresentations), a chemical language model that maps elemental composition directly to this spectrum. The encoder is pretrained by cross-modal knowledge distillation from a universal machine-learning interatomic potential; the transfer lowers the error by 11\% when only 1,000 training examples are available. On the Mat2Spec benchmark, DOSSIER reaches a mean absolute error of 3.76 states eV$^{-1}$ against 3.64 for the best structure-aware model; on an extended Materials Project dataset, the predicted spectra yield band gaps and \textit{d}-band descriptors with useful accuracy. Screening 11,977 binary and 13,251 five-component high-entropy alloy compositions for a \textit{d}-projected DOS resembling that of NiPt$_{3}$ places known oxygen reduction electrocatalysts near the top of the ranking.
\end{abstract}


\section{Main text}
\label{sec:introduction}
Electronic structure governs a broad range of properties of crystalline solids\cite{marzari2021electronic}. One of its compact representations is the electronic density of states (DOS): the number of electronic states per unit energy, obtained by integrating over the Brillouin zone\cite{martin2020electronic}. This reduction discards all \textit{k}-resolved information, yet it retains much of what determines the observable behavior of a material. In particular, the band edges set the magnitude of the band gap, and the \textit{d}-band center—the first moment of the \textit{d}-projected DOS—governs trends in adsorption energetics on transition metal surfaces\cite{hammer2000theoretical}. Experimental access to the DOS is indirect: separate spectroscopic techniques address the occupied and unoccupied states\cite{de2001high}, while the measured intensities do not reproduce the spectrum directly. Because a complete DOS profile must be assembled from several complementary and time-consuming measurements, systematic studies across many compounds rely on computed spectra, most often obtained within density functional theory (DFT)\cite{jain2011high,nair2025materials}. At the same time, the computational expense of a converged DOS calculation places an exhaustive exploration of chemical space out of reach. Machine learning surrogates\cite{kong2022density,cui2023atomic,bai2023xtal2dos,shermukhamedov2024structure,xie2025density,how2026universal} trained on DFT databases reduce this cost by orders of magnitude and reproduce the DOS from a relaxed crystal structure with an accuracy adequate for screening. However, much of the interest in materials discovery lies in underexplored chemical domains, where candidate compounds have been neither synthesized nor cataloged and their crystal structures are unknown; predicting those structures is a separate and computationally demanding problem\cite{oganov2019structure}.

Mapping composition directly to the DOS circumvents crystal structure prediction altogether. Because a single stoichiometry is compatible with a vast configurational space, such an approach has no access to long-range order and may fail to reproduce features that originate in Bloch periodicity. In contrast, the overall envelope of the spectrum is largely set by the local atomic environment, an expectation that follows from the moments\cite{cyrot1967electronic} and recursion\cite{haydock1972electronic} treatments of tight-binding theory. Composition alone does not fix the local environment, but elemental identity and stoichiometry constrain the accessible coordination numbers and oxidation states\cite{brown2009recent}. The present study examines whether the DOS can be reproduced quantitatively from the chemical composition alone (i.e., in a structure-agnostic manner), with target spectra taken predominantly from ground-state crystal structures. Two considerations support the feasibility of this task. The first is the nearsightedness of electronic matter\cite{prodan2005nearsightedness}, which suggests that the spectrum at a given site is insensitive to the arrangement of distant atoms\cite{aryanpour2025machine}. The second is the current state of composition-only models: scalar properties of inorganic materials are routinely estimated with useful accuracy in the absence of structural input\cite{alghadeer2024machine}.

A stoichiometric formula can be written as an ordered sequence of element symbols and coefficients (SI Section 2.1). This formal analogy with natural language motivates the use of chemical language models\cite{sultan2024transformers} (CLMs), transformer architectures that operate on the discrete alphabet of chemical notation rather than on ordinary text. We adopt an encoder–decoder paradigm: the encoder (i.e., CLM) compresses the tokenized composition into a latent vector, and a task-specific head then maps that vector to the target spectrum (SI Section 2.3). Here the encoder is a modernized bidirectional transformer\cite{vaswani2017attention} (ModernBERT\cite{warner2025smarter}); the decoder is a stack of transposed convolutions that progressively upsamples the latent vector into a DOS defined on a uniform energy grid. The separation of roles is intentional: the encoder has to identify which compositional patterns reflect short-range structural regularities, whereas the convolutional head supplies an inductive bias toward the local smoothness typical of DOS spectra. Following the representation-learning concept\cite{bengio2013representation}, the CLM’s initial representations of chemical formulas are formed via pretraining, before the DOS prediction task is introduced. Specifically, the model is pretrained by means of cross-modal knowledge distillation (SI Section 2.2). The teacher model is a universal machine-learning interatomic potential\cite{batatia2025foundation} trained on the potential energy surfaces of a broad range of inorganic compounds; the student CLM learns to predict its structure-derived embeddings from the chemical formula alone. Supplying structural information in this implicit way is intended to improve generalization on the downstream task while keeping inference free of crystal-structure input\cite{rubtsov2026enhancing}.

\begin{table}[b]
\caption{\textbf{Density of states (DOS) prediction performance on the Mat2Spec benchmark.} Mean absolute error (MAE) and mean squared error (MSE) are reported in states eV$^{-1}$ and states$^{2}$ eV$^{-2}$, respectively. The Density of States from Stoichiometry with Encoder Representations (DOSSIER) model with a randomly initialized encoder (knowledge distillation stage omitted) is referred to as DOSSIER$_{0}$. Numbers in parentheses denote the training subset size (e.g., 3k = 3,000; full = 30,950).}
    \begin{minipage}[b]{0.49\textwidth}
    \footnotesize
    \centering
    \setlength{\tabcolsep}{11pt}
    \renewcommand{\arraystretch}{1.3}
    \begin{tabular}{c c c c}
        \hline
        & MAE & MSE & R$^{2}$ \\
        \hline
        mean DOS & 6.39 & 138.5 & 0.15 \\
        elemental DOS & 5.58 & 113.9 & 0.32 \\
        \hline
        E3NN\cite{kong2022density} & 5.24 & 105.1 & 0.39 \\
        GATGNN\cite{kong2022density} & 4.89 & 120.9 & 0.30 \\
        Mat2Spec\cite{kong2022density} & 3.64 & 80.4 & 0.53 \\
        APET\cite{cui2023atomic} & 3.73 & 80.8 & – \\
        Xtal2DoS\cite{bai2023xtal2dos} & 3.78 & 71.2 & 0.58 \\
        EEM\cite{shermukhamedov2024structure} & 3.91 & 86.03 & – \\
        \hline
    \end{tabular}
    \label{tab:first}
    \end{minipage}
    \hfill
    \begin{minipage}[b]{0.49\textwidth}
    \footnotesize
    \centering
    \setlength{\tabcolsep}{11pt}
    \renewcommand{\arraystretch}{1.3}
    \begin{tabular}{c c c c}
        \hline
        & MAE & MSE & R$^{2}$ \\
        \hline
        DOSSIER$_{0}$ (1k) & 6.41 & 132.9 & 0.19 \\
        DOSSIER$_{0}$ (3k) & 4.92 & 102.4 & 0.39 \\
        DOSSIER$_{0}$ (10k) & 4.38 & 91.0 & 0.46 \\
        DOSSIER$_{0}$ (full) & 3.94 & 80.2 & 0.53 \\
        \hline
        DOSSIER (1k) & 5.69 & 114.5 & 0.31 \\
        DOSSIER (3k) & 4.58 & 92.2 & 0.46 \\
        DOSSIER (10k) & 4.06 & 83.7 & 0.51 \\
        DOSSIER (full) & 3.76 & 75.9 & 0.56 \\
        \hline
    \end{tabular}
    \label{tab:second}
    \end{minipage}
\end{table}

We evaluated DOSSIER (Density of States from Stoichiometry with Encoder Representations) on the Mat2Spec dataset\cite{kong2022density} (SI Section 1.1), which comprises 38,688 nonmagnetic inorganic compounds with a DOS preprocessed into 128-dimensional vectors spanning an 8 eV energy window (±4 eV relative to the band edges). Performance was quantified using the mean absolute error (MAE), mean squared error (MSE), and coefficient of determination (R$^{2}$); all results are summarized in Table 1. For comparison with published results, predicted spectra were rescaled from intensive units (states eV$^{-1}$ atom$^{-1}$) to states eV$^{-1}$ by multiplication by the number of atoms per unit cell; this quantity is available from the crystal structures in the dataset but was not accessible to the model. Two baselines are included for reference: the training-set mean spectrum and a stoichiometry-weighted combination of per-element mean spectra, where each element’s contribution is the average DOS over all training examples containing that element (SI Section 2.4). These represent successive levels of approximation—a meaningful constant predictor and a simple accounting of elemental contributions—against which the performance of composition-only models can be assessed. Metrics attained by DOSSIER on the full training subset place it on par with the top-performing structure-aware models: only Mat2Spec\cite{kong2022density} and APET\cite{cui2023atomic} yield lower MAE values, while Xtal2DoS\cite{bai2023xtal2dos} alone surpasses DOSSIER in both MSE and R$^{2}$. The value of pretraining is most apparent in the low-data regime: with 1,000 training examples, knowledge distillation reduced MAE from 6.41 to 5.69 states eV$^{-1}$ relative to random initialization—an 11\% gain that narrows to approximately 5\% on the complete training subset (3.94 vs. 3.76 states eV$^{-1}$). This trend indicates that structure-informed representations acquired during pretraining partially compensate for limited supervised signal.

\begin{figure}[b!]
  \vspace{-15pt}
  \centering
  \includegraphics[width=\textwidth]{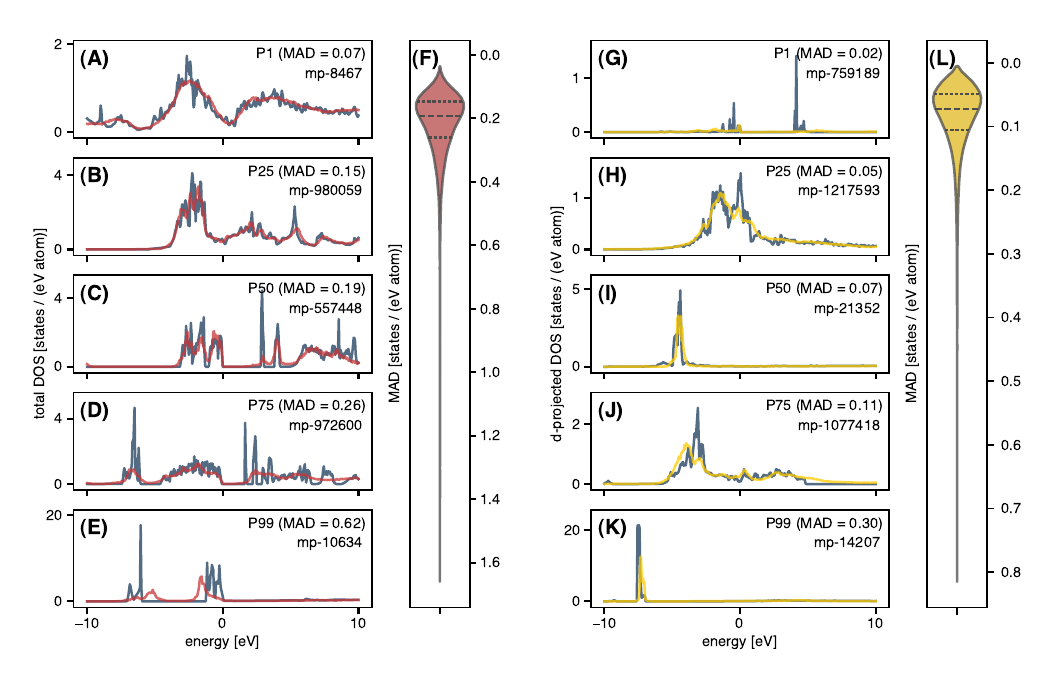}
  \caption{\textbf{Assessment of density of states (DOS) prediction on the extended Materials Project dataset.} (A–E) Total DOS at the 1st, 25th, 50th, 75th, and 99th percentiles of the per-spectrum mean absolute deviation (MAD). (F) Distribution of per-spectrum MAD for total DOS. (G–K) \textit{d}-projected DOS at the 1st, 25th, 50th, 75th, and 99th percentiles of the per-spectrum MAD. (L) Distribution of per-spectrum MAD for \textit{d}-projected DOS. All spectra are normalized per atom (states eV$^{-1}$ atom$^{-1}$); energy is referenced to the Fermi level.}
  \label{fig:fig1}
\end{figure}

To assess the fidelity of property values extracted from predicted spectra and to enable downstream high-throughput screening, we trained DOSSIER on a more recent version of the Materials Project\cite{jain2013commentary} (SI Section 1.2). This dataset includes both magnetic and nonmagnetic materials, representing broader chemical and magnetic diversity than the Mat2Spec benchmark. All spectra were interpolated onto a uniform energy grid of 401 points from –10 to +10 eV relative to the Fermi level, without removing the zero-DOS gap region. Separate models were trained for the total DOS (50,626 materials) and for the \textit{d}-projected DOS (42,996 materials); the latter is of particular relevance to heterogeneous catalysis, with the caveat that descriptors of the Hammer–Nørskov type refer to the surface-projected \textit{d}-states\cite{hammer2000theoretical}, whereas our spectra are those of the bulk. Prediction performance on the extended dataset is comparable to that observed on Mat2Spec: for the total DOS, the model achieves MAE = 3.18 ± 0.03, MSE = 84 ± 6, and R$^{2}$ = 0.47 ± 0.01 (mean ± standard deviation from 10-fold cross-validation). The corresponding metrics for the \textit{d}-projected DOS are MAE = 1.06 ± 0.01, MSE = 18.3 ± 2.2, and R$^{2}$ = 0.56 ± 0.04. Figure 1 stratifies prediction accuracy by the per-spectrum mean absolute deviation (MAD): five representative spectra at the 1st, 25th, 50th, 75th, and 99th percentiles are shown alongside the full MAD distribution. The predicted spectra closely reproduce the principal peak positions and the overall envelope of the reference up to the 75th percentile.

We next assessed the accuracy of DOS-derived property values extracted from the predicted spectra on the extended Materials Project dataset. The band gap was predicted by a dilated convolutional network parameterizing a zero-inflated lognormal distribution over band gap values (SI Section 2.5); the corresponding metrics are MAE = 0.310 ± 0.010 eV, R$^{2}$ = 0.769 ± 0.018 (Figure 2A). Metal/nonmetal classification (based on the gating probability of the same model) achieves an area under the receiver operating characteristic curve of 0.949 ± 0.004, with true positive and true negative rates of 0.89 at a metallicity probability threshold of 0.5 (Figure 2B). Analysis of prediction errors across band gap ranges (Figure 2C) shows that metallic systems (<0.05 eV) are identified with negligible error, while for nonmetals the median absolute error remains below 0.55 eV across all ranges, increasing from 0.25 eV (0.05–0.5 eV) to 0.53 eV (>5.0 eV). Four descriptors of spectral shape corresponding to the \textit{d}-projected DOS are reproduced with varying accuracy (Figure 2D–G). The center (MAE = 0.405 ± 0.009 eV, R$^{2}$ = 0.925 ± 0.006) and skewness (MAE = 0.194 ± 0.005, R$^{2}$ = 0.904 ± 0.006) are captured most precisely. The width is predicted with moderate accuracy (MAE = 0.244 ± 0.005 eV, R$^{2}$ = 0.822 ± 0.011), while kurtosis is reproduced with the lowest coefficient of determination (MAE = 0.675 ± 0.023, R$^{2}$ = 0.723 ± 0.025).

\begin{figure}[b!]
  \vspace{-15pt}
  \centering
  \includegraphics[width=\textwidth]{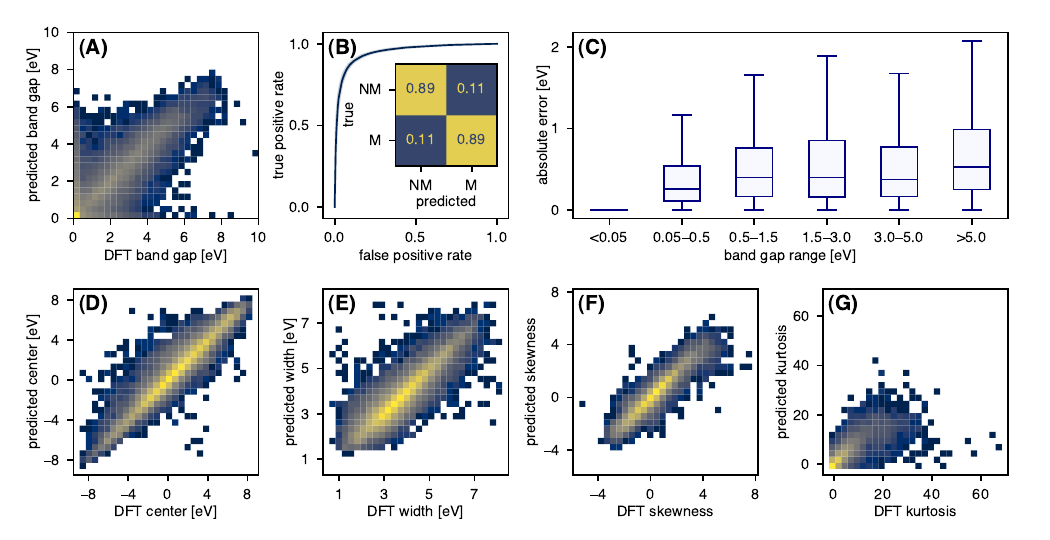}
  \caption{\textbf{Accuracy of property values derived from predicted density of states (DOS) on the extended Materials Project dataset.} (A) Predicted and density functional theory (DFT) band gap values. (B) Receiver operating characteristic curve for metal/nonmetal (M/NM) classification; inset: confusion matrix at a metallicity probability threshold of 0.5 (values normalized by true class). (C) Distribution of absolute band gap errors across band gap ranges. (D–G) Predicted and DFT values of the first four spectral moments of the \textit{d}-projected DOS.}
  \label{fig:fig2}
\end{figure}

To demonstrate the practical utility of the trained models, we performed high-throughput screening of binary and high-entropy alloy (HEA) candidates, ranking each composition by the similarity of the predicted \textit{d}-projected DOS to that of NiPt$_{3}$, a prototypical oxygen reduction reaction (ORR) catalyst\cite{stamenkovic2007improved} (SI Section 3.1). The binary alloy pool comprised 11,977 compositions A$_{m}$B$_{n}$ ($m, n \le 3$) enumerated from metallic elements (SI Section 3.2). For each candidate, the \textit{d}-projected DOS was predicted by the 10-model ensemble and scored by a consensus metric combining cosine similarity and the first Wasserstein distance, both computed in the catalytically relevant window from –8 to +2 eV relative to the Fermi level (SI Section 3.3). The top-ranked candidate is NiPt$_{3}$ itself, confirming that the model reproduces the reference spectrum with high fidelity. Excluding the reference composition, the three highest-ranked binary candidates are Rh$_{3}$Au, NiIr$_{2}$, and NiPt$_{2}$ (Figure 3). Bang et al.\cite{bang2024inverse} applied an inverse design model to the same catalytic target and reported five binary candidates: NiPt$_{3}$, CoPt$_{3}$, FePt$_{3}$, Mo$_{3}$Ni, and Mo$_{3}$Co. The two studies differ in spectral representation (total DOS from –7.5 eV to +7.5 eV in ref.\cite{bang2024inverse}), but all five compositions fall within the top 2\% of our ranking (1st, 9th, 25th, 174th, and 126th, respectively). This overlap is consistent with the dominance of the \textit{d}-projected component in the total DOS within the considered energy window (Figure 3).

The same protocol was applied to five-component equiatomic HEAs: 13,251 compositions passed the phase-stability filter (SI Section 3.2) and were ranked by the consensus score. The ten highest-ranked HEA candidates are shown in Figure 3; the most common elements are Au (nine occurrences), Ni (eight), Ir (seven), Pt and Co (five each), and Ru, Rh, and Cu (four each). Among the HEA compositions previously reported as ORR catalysts, RuRhPdIrPt\cite{chen2026high} ranks 193rd, CoPdNiCuPt\cite{nevelskaya2025high} 373rd, and RuCuOsIrPt\cite{chen2015multi} 455th, all within the top 5\% of the pool. This correspondence suggests that d-projected DOS similarity can serve as a screening descriptor for ORR catalyst candidates. Two further reported catalysts are ranked lower: FeCoPdNiPt\cite{yu2022high} at 914th and FeCoNiCuPt\cite{chen2023ptfeconicu} at 1897th, corresponding to the top 7\% and top 14\%, respectively. Compositions ranked relatively low are those with electronic structures unlike that of NiPt3, and their activity is presumably governed by factors outside the considered descriptor.

A fundamental limitation of the composition-only approach is its inability to distinguish polymorphs: a single stoichiometric formula may correspond to several crystal structures whose electronic spectra differ, yet the absence of structural input restricts the output to one spectrum per composition. On the Mat2Spec benchmark, DOSSIER matches structure-aware models—a result that reflects the properties of the dataset rather than showing that polymorphism has been handled. Specifically, 38,688 structures span 31,873 unique compositions; 27,925 (88\%) of these are represented by a single polymorph. Most compounds are located near the convex hull: 41\%, 71\%, and 81\% have energies above the hull not exceeding 0, 50, and 100 meV per atom, respectively. Therefore, the Mat2Spec benchmark affords limited insight into the impact of polymorphism on prediction accuracy.

\begin{figure}[b!]
  \vspace{-15pt}
  \centering
  \includegraphics[width=\textwidth]{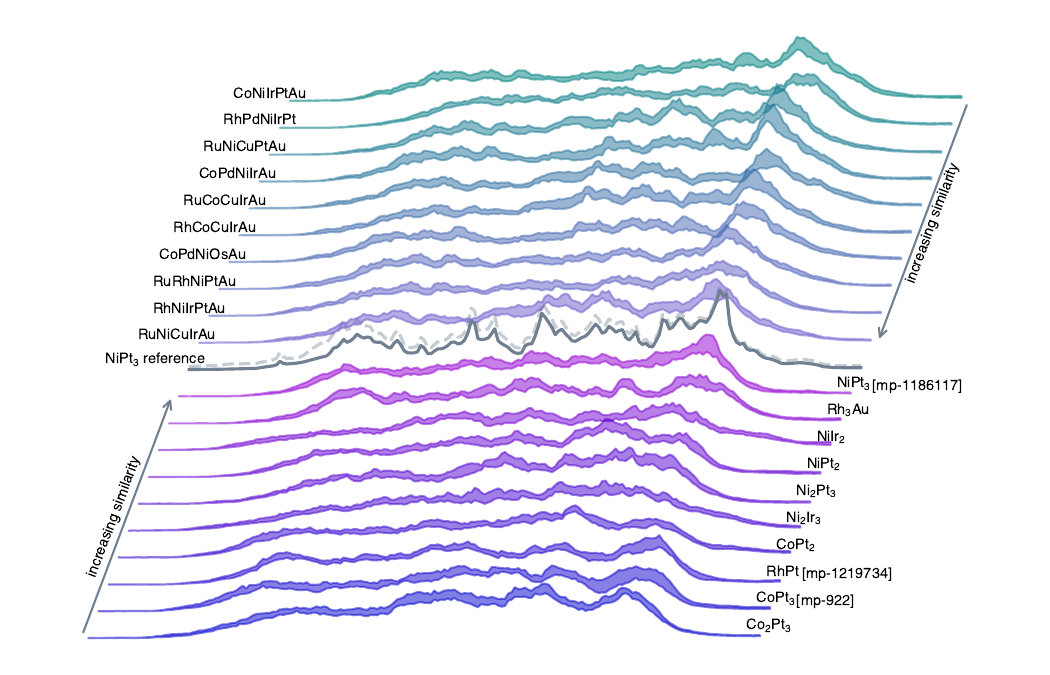}
  \caption{\textbf{High-throughput screening for oxygen reduction catalysis: top-ranked binary and high-entropy alloy (HEA) candidates.} Compositions are ranked by \textit{d}-projected density of states (DOS) similarity to the NiPt$_{3}$ reference using the consensus score. The gray solid and dashed lines show the NiPt$_{3}$ reference \textit{d}-projected DOS and total DOS, respectively. The ten highest-ranked binary alloys are shown below the reference, and the ten highest-ranked five-component equiatomic HEAs above. In both groups, spectra are ordered by increasing similarity to the reference. Uncertainty envelopes correspond to ±1 standard deviation across the 10-model ensemble. All spectra are normalized per atom (states eV$^{-1}$ atom$^{-1}$).}
  \label{fig:fig3}
\end{figure}

A more detailed analysis was carried out on the extended Materials Project dataset, where 6,958 compositions possess metastable configurations in addition to the ground-state structure, yielding 12,324 metastable polymorphs that were not used for model development. All spectra in this subset were normalized per atom to account for differing unit-cell sizes; all values are in states eV$^{-1}$ atom$^{-1}$. The MAE between the model predictions and the DOS spectra of the corresponding stable polymorphs is 0.209 ± 0.007. The MAE between the same predictions and the DOS spectra of the metastable polymorphs increases to 0.233 ± 0.010, as expected given that the model was trained on ground-state structures exclusively. The comparison between the DOS spectra of stable and metastable polymorphs themselves yields a mean absolute deviation of 0.258 ± 0.010; the same quantity between randomly paired materials across the entire collection is 0.526. Polymorphism is therefore a substantial source of variation at fixed composition. For metastable phases, the model’s prediction error is bounded by the intrinsic polymorph-to-polymorph divergence, indicating that further improvement would require structural input.

The application of the presented models to HEA screening extends beyond their formal domain of applicability. At the same time, direct DFT screening of HEA candidates is computationally prohibitive; a single equiatomic quinary composition admits numerous surface configurations. The composition-to-DOS mapping, despite its polymorphism limitation, provides one of the few tractable routes to electronic-structure-informed screening across a diverse compositional space.

Despite these limitations, our study explicitly demonstrates the practical value of models that directly map chemical composition to the DOS, a compact representation of the electronic structure of inorganic compounds. The presented approach, based on a chemical language model, complements structure-aware methods and DFT calculations that supply their training data, offering a first-level screening tool for identifying promising candidates in underexplored chemical domains.

\section{Conflicts of interest}
\label{sec:conflicts}
There are no conflicts of interest to declare.

\section{Acknowledgements}
\label{sec:acknowledgements}
This work was supported by the Ministry of Economic Development of the Russian Federation in accordance with the subsidy agreement (agreement identifier 000000C313925P4H0002; grant No 139-15-2025-012).

\section{Data availability statement}
\label{sec:data_availability}
The Mat2Spec benchmark dataset is publicly available from the official repository at \url{https://github.com/gomes-lab/Mat2Spec}, while the extended Materials Project dataset was compiled using the official Materials Project API (\url{https://materialsproject.org/api}). An interactive online predictor for the total density of states (DOS) is freely accessible on HuggingFace Spaces at \url{https://huggingface.co/spaces/zxcghoul3228/DOSSIER_demo}.

\bibliographystyle{unsrt}
\bibliography{references}

\end{document}